\documentstyle[prl,aps]{revtex}
\begin{document}
\draft
\title{Topologically Massive Gravity and Black Holes in Three Dimensions}
\author{T. Dereli \, , \,  \"{O}. Sar{\i}o\u{g}lu}
\address{ Department of Physics ,  Middle East Technical University, 
06531 Ankara, Turkey}

\maketitle

\bigskip

\begin{abstract} 
We obtain a general class of exact
solutions to topologically massive gravity
with or without a negative cosmological constant.
In the first case, we show that the solution is supersymmetric
and asymptotically approaches
the extremal BTZ black hole solution, 
while in the latter case it goes to flat space-time.
\end{abstract}

\bigskip


\bigskip
 
The discovery of  BTZ black holes \cite{BTZ}
enhanced the interest in (1+2) dimensional gravity models 
considerably as their thermodynamic properties
closely resemble those of (1+3) dimensional black holes. However, it is
well known that general relativity in (1+2) dimensions
has no propagating degrees of freedom and no
Newtonian limit (see e.g. \cite{carlip} and the references therein).
A physically interesting modification of the (1+2) dimensional general
relativity that cures at least some of these deficiencies
is provided by the addition of the gravitational Chern-Simons
term to the usual Einstein-Hilbert term in the action. 
This theory is usually called 
topologically massive gravity (TMG) \cite{djt}, whose
field equations include the
Cotton tensor, which is the analogue of the Weyl tensor in three dimensions, 
in addition  to the usual Einstein tensor. With this addition
new degrees of freedom are introduced
and one now has a dynamical theory with a massive
graviton. 
The BTZ metric satisfies the TMG field equations in a trivial way as the 
Cotton tensor vanishes identically.
Black hole type solutions for which the Cotton tensor is not trivial 
has been long sought for.
In fact a solution to the linearized version of TMG for a stationary
rotationally symmetric source was found and 
it was conjectured that there are no asymptotically flat stationary
solutions in the absence of any sources \cite{des}.
Here we exhibit a new class of exact supersymmetric solutions 
that, with an appropriate choice of integration
constants, may have event horizons
and asymptotically approach
the extremal BTZ black hole solution.

We consider the action $ I[e,\omega]  = \int_M L$ where
the Lagrangian 3-form
\begin{equation}
L= \frac{1}{\mu} (\omega^{a}_{\;b} \wedge  d \omega^{b}_{\;a} + \frac{2}{3} 
\omega^{a}_{\;c} \wedge \omega^{c}_{\;b} \wedge \omega^{b}_{\;a}) +
\frac{1}{2} {\cal R} \; *1 - \lambda \; *1 \;\; ,
 \;\; \label{act}
\end{equation}
contains the Einstein-Hilbert term, 
a negative cosmological constant
$\lambda = - 1/l^2 < 0$ and the gravitational Chern-Simons term with the
coupling constant $\mu$, written in terms of 
Levi-Civita connection 1-forms $\omega^{a}_{\; b}$.
Thus the variation of $I$ with respect to orthonormal coframes $e^a$ yields:
\begin{equation}
\frac{1}{\mu} \; C_a + G_a + \lambda \; *e_a  = 0 \;\; , \label{fieq}
\end{equation}
where the Einstein 2-forms \( G_a \equiv G_{ab} *e^b = 
- \frac{1}{2} R^{bc} \; *e_{abc} \)
and the Cotton 2-forms  \( C_a \equiv D Y_a = d Y_a + 
\omega_a^{\;b} \wedge Y_b . \) 
We defined
\( Y_a \equiv (Ric)_a - \frac{1}{4} {\cal R} e_a \),
in terms of the Ricci 1-forms 
\( (Ric)_b \equiv \iota_a \; R^a_{\;b} \),
and the curvature scalar
\( {\cal R} \equiv \iota_a \; (Ric)^a \). 
Here \( R^a_{\;b} = d \omega^a_{\;b} + \omega^a_{\;c} \wedge \omega^c_{\;b} \)
are the curvature 2-forms of the 
Levi-Civita connection 1-forms
that satisfy  Cartan structure equations
\( d e^a + \omega^a_{\;b} \wedge e^b = 0. \)  
Hodge duality is specified by the oriented volume element
$*1 = e^0 \wedge e^1 \wedge e^2.$

The solutions will be given in terms of
 the local coordinates $(t,\rho,\phi)$ 
by the metric tensor
\begin{equation}
g = - e^0 \otimes e^0 + e^1 \otimes e^1 + e^2 \otimes e^2 \;\; , \label{met}
\end{equation}
where we choose
\begin{equation}
e^0 = f(\rho) dt \;\; , \;\; e^1 =  d\rho \;\;, \;\; 
e^2 = h(\rho) ( d \phi + a(\rho) dt ). \;\; \label{co1fo}
\end{equation}
 Denoting the derivatives with respect to 
 $\rho$ by a prime, the connection
1-forms are found to be:
\begin{equation}
w^0_{\;1} = \alpha e^0 - \frac{1}{2} \beta e^2 \;\;  ,  \;\;
w^0_{\;2} = - \frac{1}{2} \beta e^1 \;\;  ,  \;\;
w^1_{\;2} = - \frac{1}{2} \beta e^0 - \gamma e^2 \;\;  ,  \label{con1fo}
\end{equation}
where we set the connection coefficients
\begin{equation}
\alpha \equiv \frac{f^\prime}{f} \;\; , \;\;
\beta \equiv \frac{a^\prime \, h}{f} \;\; , \;\;
\gamma \equiv \frac{h^\prime}{h} \;\; .  \label{albega} 
\end{equation}
The corresponding curvature 2-forms turn out to be
\begin{equation}
R^0_{\;1} = A e^1 \wedge e^0 + B e^2 \wedge e^1 \;\;, \;\;
R^0_{\;2} = C e^2 \wedge e^0  \;\;, \;\;
R^1_{\;2} = B e^0 \wedge e^1 + D e^2 \wedge e^1 \;\; , \label{cur2fo} 
\end{equation}
where we defined 
\[ A \equiv \alpha^\prime + \alpha^2 - \frac{3}{4} \beta^2 \;\; , \;\;
B \equiv \frac{1}{2} \beta^\prime + \gamma \, \beta \;\; , \;\;
C \equiv \alpha \, \gamma + \frac{1}{4} \beta^2  \;\; , \;\; 
D \equiv \gamma^\prime + \gamma^2 + \frac{1}{4} \beta^2 .\]
After some algebra the field equations can be reduced to:
\begin{eqnarray}
-D + \frac{1}{l^2} + \frac{1}{\mu} [B^\prime + B \gamma + \frac{1}{2}
\beta (C-A)] 
& = & 0 \;\; , \label{tmg1} \\
-B + \frac{1}{\mu} [\frac{1}{2} (D-A-C)^\prime + \alpha (D-C) + 
\frac{3}{2} \beta B] & = & 0 \;\; , \label{tmg2} \\
C - \frac{1}{l^2} + \frac{1}{\mu} [(\gamma - \alpha) B + \frac{1}{2} \beta
(A-D)] 
& = & 0 \;\; , \label{tmg3} \\
A - \frac{1}{l^2} + \frac{1}{\mu} [B^\prime +\alpha B + \frac{1}{2} \beta
(C+D-2A)] 
& = & 0 \;\; . \label{tmg4} 
\end{eqnarray}

We found it remarkable that 
the following conditions on the connection coefficients 
\begin{equation}
\alpha = \frac{k}{2} \beta + \frac{1}{l}  \;\; , \;\;
\gamma = - \frac{k}{2} \beta + \frac{1}{l} \;\; , \label{alga}
\end{equation}
(with $k^2=1$), 
that follow from the field equations and 
were essential for finding the general
self-dual solutions of the Einstein-Maxwell-Chern-Simons theory in
(1+2) dimensions \cite{Dereli},
turn out to yield solutions in the present case as well.
In fact, the conditions (\ref{alga}) have a significance in the following
sense: Any solution of topologically massive gravity is said to be
supersymmetry preserving provided there exists a nontrivial real 2-spinor
$\epsilon$ satisfying \cite{sugra}
\begin{equation}
(2 {\cal D} + \frac{1}{l} \gamma ) \epsilon = 0 \;\;\; \label{susy}
\end{equation}
where \( \gamma = \gamma_a e^a \) and \( {\cal D} = d + \frac{1}{2}
w^{ab} \, \sigma_{ab} \) with \( \sigma_{ab} = \frac{1}{4}
[\gamma_a,\gamma_b] \) .
With a suitable choice of real $\gamma$-matrices and after some algebra
we find that (\ref{alga}) are in fact the necessary and sufficient
conditions for supersymmetry. It should be noted, however, that 
with the above assumption only the extremal BTZ solution can be
recovered as $|\mu| \rightarrow \infty.$

The curvature components are now 
greatly simplified  and they are given by
\begin{equation}
A = U + \frac{1}{l^2} \;\; , \;\; B = kU \;\; , \;\; C = \frac{1}{l^2}
\;\; , \;\;
D = \frac{1}{l^2} - U \;\; \label{ABCD}
\end{equation} 
where 
\begin{equation} 
U = - \frac{1}{2} \beta^2 + \frac{k}{l} \beta + \frac{k}{2} \beta^\prime
\;\; .
\label{defu}
\end{equation}
The equations (\ref{tmg1}) - (\ref{tmg4}) are satisfied simultaneously
provided $U$  satisfies
\begin{equation}
U^\prime - k \beta U + (\frac{1}{l}+k \mu) U = 0 \;\; . \label{eqnu}
\end{equation}
By setting $V = k \frac{\beta}{U}$ in (\ref{eqnu}),  we arrive at the 
linear first order
ordinary differential equation
\begin{equation}
V^\prime + (\frac{1}{l}-k \mu) V = 2 \;\; . \label{eqnv}
\end{equation}
This is easily integrated and 
\[ V = \frac{2}{1/l -k \mu} [1+ \beta_0 e^{-(1/l - k \mu) \rho}] \]
for some integration constant $\beta_0$. Hence, going back to the 
definition of
$U$ (\ref{defu}) and substituting for $V$, we obtain a differential
equation for $\beta$ as:
\begin{equation}
\beta^\prime + \beta \left(\frac{2}{l} - 
\frac{(1/l -k \mu)}{1+ \beta_0 e^{-(1/l -k \mu) \rho}} \right) - k \beta^2
= 0 
\;\; . \label{eqnbeta}
\end{equation}
Setting  $\omega = 1/\beta$, one finds:
\begin{equation} 
\omega^\prime +  \left(
\frac{(1/l -k \mu)}{1+ \beta_0 e^{-(1/l -k \mu) \rho}} - \frac{2}{l}
\right)\omega 
+ k = 0 \;\; . \label{eqnom}
\end{equation} 
When integrated, this yields:
\begin{equation}
\beta = \frac{1}{\omega} = k \;
\frac{2/l + \beta_2 (1/l +k \mu) e^{(1/l -k \mu) \rho}}
{1+\beta_1 e^{2 \rho /l}+ \beta_2 e^{(1/l -k \mu) \rho}}
\;\; \label{solbe}
\end{equation}
for integration constants 
\( \beta_2 \equiv \frac{2}{l \beta_0 (1/l+ k \mu)} \) and $\beta_1$. 
Finally the metric functions are found to be
\begin{eqnarray}
f & = & f_0 e^{2 \rho / l} [1+\beta_1 e^{2 \rho /l}+ 
\beta_2 e^{(1/l -k \mu) \rho} ]^{-1/2} \;\; , \label{solf} \\
h & = & h_0 [1+\beta_1 e^{2 \rho /l}+ 
\beta_2 e^{(1/l -k \mu) \rho}]^{1/2} \;\; , \label{solh} \\
a & = & -a_0 + k \; \frac{f_0}{h_0} e^{2 \rho /l} 
[1+\beta_1 e^{2 \rho /l}+ \beta_2 e^{(1/l -k \mu) \rho}]^{-1} \;\; ,
\label{sola}
\end{eqnarray}
where $a_0$, $f_0$ and $h_0$ are some new integration constants.


Depending on the values of the integration
constants $\beta_1$ and $\beta_2$ (of course as well as on $l$ and $\mu$),
one might have singularities in these metric functions. 
It is not difficult to verify that   for $1+\beta_2 > \beta_1 > 0$, 
the metric function $g_{tt}$ changes sign for some $\rho_0 \in [0, \infty)$. 
 However, an analysis as the one given in \cite{BTZ} 
cannot be given here since,
at the very starting point, it is impossible for one to invert the
functional relation \( r = h(\rho) \) and to rewrite the metric
 in terms of $r$. That step is crucial for one to
convert the metric into the well studied form of the BTZ (and hence the
AdS) metric and make use of the vast literature on that subject.
Then we ask what else can be done and for that we go
back to the full solution and analyze the quasilocal
mass and the angular momentum. We refer the reader to \cite{Dereli} for
a discussion of how these quantities can be found in this AdS background.
For the quasilocal angular momentum, we have 
\begin{equation}
j(r)=k h_0^2 \varphi(r)
\end{equation}
where \( \varphi(\rho) \equiv 2/l + \beta_2 (1/l+k \mu) e^{(1/l-k
\mu) \rho} \)
and again one finds that one has to invert
$r=h(\rho)$ so that $\varphi$ can be written as
a function of $r$. Similarly the quasilocal energy turns out to be
\begin{equation}
E(r)=\frac{h_0^2}{2 r} \varphi(r) = \frac{k}{2 r} j(r) 
\end{equation}
whereas the quasilocal mass is
\begin{equation}
m(r)= a_0 j(r) = k a_0 h_0^2 \varphi(r) \;\; .
\end{equation}
The total angular momentum $J$ and the total mass $M$ are defined by
the limits \( J \equiv j(r) |_{r \rightarrow \infty} \) and
\( M \equiv m(r) |_{r \rightarrow \infty} \), respectively. To see
what can be said about $J$ and $M$, we first start by examining $a(r)$.
Depending on the values of $l$ and $\mu$, $a$ either goes to $-a_0$ or
$-a_0+ \frac{k f_0}{h_0 \beta_1}$ as $r \rightarrow \infty$. Hence for
$a$ to vanish asymptotically as $r \rightarrow \infty$, $a_0$ should
be chosen either as 0 or as $\frac{k f_0}{h_0 \beta_1}$. When $1/l > k\mu$,
$a_0=0$ and hence $M=0$ whereas $J \rightarrow \infty$. For $1/l <k\mu$,
$J$ is finite and $J=2 k h_0^2 /l$. Then $a_0 = \frac{k f_0}{h_0 \beta_1}$
and $M=a_0 J$ is finite as well. 
As $k\mu \rightarrow \infty$, this solution approaches the extremal BTZ
solution. To see this, use the freedom to choose the radial coordinate
and replace $\rho$ by $r=h(\rho)$. So now 
$e^1=g(r) \, dr$ for some function $g$ and
\[ g \equiv \frac{d \rho}{dr} = \frac{l r}{(r^2-h_0^2)} \;\; ,
\;
f = \frac{f_0 h_0}{\beta_1} \left( \frac{r}{h_0^2} - \frac{1}{r} \right) 
\;\; , \; a = -a_0 + \frac{k f_0}{h_0 \beta_1} 
\left( 1 - \frac{h_0^2}{r^2} \right) \;\; . \]
Comparing with the BTZ solution \cite{BTZ}, \cite{carlip}, it is easily
seen that choosing \( J=2 h_0^2/l \), \( M=J/l \), $k=1$, $a_0 =1/l$ and
\( \frac{f_0}{\beta_1} = h_0/l  \), one gets the extremal BTZ solution.
It is well known that the BTZ solution is quite similar to the Kerr
solution in 3+1 dimensions \cite{BTZ}. Since both our solution 
and the extremal BTZ solution are supersymmetric \cite{sugra} (just
like the extremal Kerr solution in 3+1 dimensions), it is perhaps not
surprising that one gets the extremal BTZ solution in the limit.

In the absence of a cosmological constant, the solution
has to be reanalyzed since simply setting $1/l = 0$ in the above 
expressions does not give the desired limiting solution. In this 
case, $V$ is now
\begin{equation}
V = \frac{-2}{k \mu} (1+\beta_0 e^{k \mu \rho}) \;\; , \label{newsolV}
\end{equation}
whereas the equation for $\omega$ becomes
\begin{equation}
\omega^\prime - \frac{k \mu}{1+ \beta_0 e^{k \mu \rho}} \omega + k = 0 \;\; .
\label{neweqnom}
\end{equation}
Integrating this, one finds:
\begin{equation}
\beta = \frac{1}{\omega} = \frac{\mu (1+ \beta_0 e^{k \mu \rho})}
{1-\mu \beta_0 (\omega_0+k \rho) e^{k \mu \rho}} \;\; \label{newsolbe}
\end{equation}
for some integration constant $\omega_0$. The new metric functions are
finally found to be:
\begin{eqnarray}
f & = & f_0 [e^{-k \mu \rho} - \mu \beta_0 (\omega_0 +k \rho)]^{-1/2} \;\; , \\
h & = & h_0 [e^{-k \mu \rho} - \mu \beta_0 (\omega_0 +k \rho)]^{1/2} \;\; , \\
a & = & -a_0 +k \frac{f_0}{h_0} 
[e^{-k \mu \rho} - \mu \beta_0 (\omega_0 +k \rho)]^{-1} \;\; , 
\end{eqnarray}
where $f_0$, $h_0$ and $a_0$ again denote some integration constants.
For this case, all the nontrivial components of the curvature 2-forms
$R^a_{\;b}$ are proportional to $U$, as can be easily seen by examining
(\ref{cur2fo}) and (\ref{ABCD}). By using (\ref{newsolV}) and (\ref{newsolbe})
in the definition of $U = k \frac{\beta}{V}$, we find that
\begin{equation}
U = - \frac{\mu^2}{2} \; 
\frac{1}{1-\mu \beta_0 (\omega_0+k \rho) e^{k \mu \rho}} \;\; . 
\label{newsolU}
\end{equation}
If $k \mu > 0$, then as $\rho \rightarrow \infty$, $U \rightarrow 0$ and
hence $R^a_{\;b} \rightarrow 0$, which implies that this solution 
asymptotically approaches flat space. 
However, we checked that it is not accesible to linearization
about flat space in the sense of Ref. [4].

\bigskip

In summary, we have obtained a new solution to the topologically massive
gravity model with a negative cosmological constant. We have shown that it 
asymptotically approaches the extremal BTZ solution, and depending on the
integration constants, has event horizons. Moreover it does go to 
flat space as one sets the cosmological constant to zero. 

\bigskip

We thank Professors S. Deser and Yu. N. Obukhov for 
enlightening discussions.

\end{document}